\documentclass{IEEEconf}
\usepackage{url}
\usepackage[latin9]{inputenc}

\author{
  Samuel Thibault\\
  \begin{affiliation}
    LaBRI\\
    351 cours de la Libération\\
    33405 Talence Cedex\\
    France
  \end{affiliation} \\
  \email{Samuel.Thibault@ens-lyon.org}\\
  Tel: +33 5 40 00 35 40 Fax: +33 5 40 00 66 69
\and
  Sébastien Hinderer\\
  \begin{affiliation}
    LORIA\\
    615 rue du Jardin Botanique\\
    54600 Villers-lès-Nancy\\
    France
  \end{affiliation} \\
  \email{Sebastien.Hinderer@ens-lyon.org}\\
  Tel: +33 3 83 59 20 35 Fax: +33 3 83 27 83 19
}
\title{
BrlAPI: Simple, Portable, Concurrent, \\
Application-level Control of Braille Terminals}

\begin{document}
\maketitle

\begin{abstract}
Screen readers can drive braille devices for allowing visually impaired
users to access computer environments, by providing them the same
information as sighted users.  But in some cases, this view is not
easy to use on a braille device.  In such cases, it would be
much more useful to let applications provide their own braille feedback,
specially adapted to visually impaired users.  Such applications would
then need the ability to output braille ; however, allowing both screen
readers and applications access a wide panel of braille devices is not a
trivial task.

We present an abstraction layer that applications may use to communicate
with braille devices.  They do not need to deal with
the specificities of each device, but can do so if necessary.
We show how several applications can communicate with one braille
device concurrently, with BrlAPI making sensible choices about which
application eventually gets access to the device. The description of
a widely used implementation of BrlAPI is included.
\end{abstract}

\section{Introduction}

Usually, visually impaired people use computers with the help of special
programs called screen readers which deliver the information
displayed on the screen using a speech synthesis system or a refreshable
braille display. Designing screen readers able to communicate with
different brands of braille display was so difficult that until recently
screen readers could only work with one brand of braille terminal. It was
then usual for a manufacturer to provide both a braille terminal and a
screen reader using it. Only the most recent screen readers support
several brands of braille terminals.

However, even with these newer readers, there is a deeper problem that
remains unsolved: visually impaired people are forced to use exactly
the same interface as sighted people, since there is no standard
mechanism for an
application to provide an alternative, more suited interface in addition
to the standard screen-based one. One part of this problem lies in the
fact that there is no way for applications to communicate directly with
braille terminals, since these are being used exclusively by screen readers.

The BrlAPI framework that will be introduced in this paper can be seen
as a first step towards the resolution of this problem. Indeed, it
proposes a client-server approach that lets applications communicate with
braille terminals thanks to a server that both removes the need for
applications to know exactly how to communicate with each braille
terminal and lets several applications share the braille device.
Device-specific applications like file transfer tools are not neglected either,
since BrlAPI offers a way for such programs to communicate directly
with the device.

In Section~\ref{problems} we discuss in greater detail the problems that
are specific to braille display. Section~\ref{solution} describes the
BrlAPI solution; an implementation of this solution is presented in
Section~\ref{implementation}. Finally, Section~\ref{results} presents
the current clients of BrlAPI, and Section~\ref{conclusion} gives some
concluding remarks and lists a few remaining open questions.

\section{Issues for Driving Braille Devices} \label{problems}

The question of accessing a braille display within an application is not
trivial, mainly because of the large variety of available models,
and because of the problem of properly sharing
one display between several
applications. Protecting the access to braille
displays must also be considered.

\subsection{Heterogeneity of Braille Devices} \label{variety}

Every braille terminal has a
refreshable braille display divided into eight pin cells; however, the number of available
cells and their layout vary from display to display. Some terminals have
a unique display, others have two (the main one, generally used to display a
small region of the screen, and an auxiliary one, usually used to report
status information like cursor position or current time). Some modern
devices also let users adjust the firmness of braille dots to either
display hard and easy-to-read dots when the display is connected to a
power source, or softer
dots when the display is unplugged, to save energy.

Perhaps more importantly, braille terminals come with very different
keyboards. Some terminals which are also used as note-takers include a
keyboard to write characters (either in braille or through a more
standard PC-like keyboard). In addition to these optional
keyboards, braille terminals have function keys whose number and layout
is, again, terminal dependent. Manufacturers of braille terminals
have shown a rather vivid imagination regarding the kind of available
function keys: simple keys, keys associated with braille cells (usually
called routing keys because one of their functions is to bring the
cursor to the character displayed by their associated cell), joysticks,
navigation bars that can be moved one or two steps in each of four
directions, wheels, {\em etc.} To sum up, braille keyboards can be
much more complex than plain PC keyboards. In fact, they often provide
the functionality available both through a keyboard and a mouse.

Last but not least, there is no standard way to communicate with a
braille display.
Each manufacturer has designed its own communication protocol to let its
braille devices receive text to display or configuration information
and to send keyboard events or report status information.

Hence, the first problem one encounters is that braille terminals form a
completely heterogeneous class of devices. How should displays be
accessed? How should keyboard events be delivered? Should they be
rather driver specific (each application then has to interpret them
in a consistent manner), or should keyboard events be described in a
more standard way, in spite of their potentially different nature ?

\subsection{Concurrency} \label{concurrency}

Another issue is that modern operating systems let several
applications run concurrently, their visual output being usually
sent to separate virtual terminals or windows.  Some of these applications
may want to use the braille device for providing more appropriate braille
output.  A visually impaired user may also want to use several screen readers
simultaneously, in order to get the best benefits from each of them.
This results in a concurrency issue: what should eventually be
displayed on the braille terminal?  To which application should the key
events be directed?

Moreover, one may want to write some applications dedicated to a given
brand of terminals, that would take full control of the braille device.
For instance, note-takers can often exchange files with the
computer, through a brand-specific protocol.  So as to avoid corruption
during the transfer, all other device operations should be suspended,
but the applications that requested them should not have to be aware of that.

\subsection{Protection} \label{protection}

Usually, input/output devices such as keyboards and screens are
controlled by the operating system's kernel.
Each user process has to
interact with the kernel to perform input/output. Therefore, it is unlikely that a user
process can access these devices in a way which would render them unusable.
This, however, is not true for braille devices.
Indeed, they are generally handled by a user-mode process, since this
makes development and distribution easier. As a consequence, braille
terminals are much more vulnerable than more traditional input/output
devices to incorrect use by other processes. Given that incorrect
use can damage a braille terminal irreversibly, and  that these devices are
extremely expensive, it is important to ensure good
protection of this resource against incorrect usage or malicious
intent. This need for good protection becomes even more critical
in the context of modern multi-user networked operating systems.

\section{Proposed Solution} \label{solution}

For tackling all the issues described in the previous Section, we
propose the BrlAPI framework, based on a client-server approach.
A BrlAPI server uses a driver for controlling the braille terminal, and
applications connect to the server for interacting with the braille
device. This even allows an application running on one machine to drive
a braille display connected to another machine, provided that the two
machines are interconnected by a network link.

\subsection{Handling the Variety of Terminals} \label{variety_handle}

The BrlAPI server has a series of drivers, each of them knowing how to
communicate with one given family of braille devices. More precisely,
the server is able to send text to display, and to receive key events.
Optionally, a driver can also provide lower-level, packet-based
input/output functionality. In
practice, one running instance of the BrlAPI server communicates with {\em
one} braille device.

When clients connect to a server, they can request the dimensions of the
corresponding display, and then perform \verb!write! requests.

Moreover, clients can query the server for the name of the braille
device it is communicating with. This permits a client to decide how it
wants to receive key events. Key events can be delivered
to a client application in two different ways.

First, the server can deliver key events it gets from the device as-is.
This presupposes that client applications know exactly how key events are
returned by the braille driver they are talking to, and lets developers
use the knowledge of some given keyboard when designing their
applications.

Second, key events can be converted to a standard representation by
the server before being delivered to a client. For instance, braille
devices often have keys typically used for browsing the current line,
going to the next or previous line, {\em etc.}  Clients can then be written
independently of the braille device, their use remaining easy and
intuitive. This also favours a semantically consistent use of keys
from one client application to another.

\subsection{Selecting the Proper Application}
\label{concurrency_handle}

Previous work like Libbraille \cite{libbraille} achieved results similar
to those described in Section \ref{variety_handle}.  However, the context of
Libbraille was the TIM project \cite{tim}, in which only one application
needed to be run at a time.  As a result, Libbraille was intrinsically
not designed to allow several applications to control the braille
terminal.

In contrast, thanks to the client-server design, several
applications can connect to the BrlAPI server.  Each application
behaves like if it were the only one that has access to the braille device,
but the BrlAPI server actually \emph{switches} between applications
as appropriate.  Only one application at a time has the ``focus'',
which means that its output is actually sent to the braille device,
and braille key events are sent back to it.  Output of each other
application is stored and will be displayed later, when
they get the focus again.

What ``appropriate focus'' may mean is not completely clear.  One natural way of
switching between applications is to just follow the already well-known
keyboard and mouse focus.  This is necessary when using braille
terminals which do not have an integrated keyboard, since the regular
keyboard then has to be used.  Nevertheless, even when using a braille
terminal with an integrated keyboard, it is a lot easier for a visually
impaired
person to work with a sighted person if the notion of \emph{focus} is
the same for both of them.  It should be noted that this is different from
the situation of dispatching speech synthesis \cite{speechd}. The reason
for this is that speech naturally expresses \emph{time}, while braille
naturally expresses \emph{space}.

The solution we adopted is hence to require applications, before issuing
write requests or starting to wait for key events, to declare ``where''
they are running, this being expressed as an integer, and the BrlAPI server just needs
to find out where the keyboard and mouse focus is. In the case of the Linux
console for instance, this is the number of the Virtual Terminal (VT) in
which the application is running, and the BrlAPI server just asks the kernel
which VT is active.  In the case of an X-Window desktop, this is the ID
of the window in which the application is running, and a dedicated X-window
application called \texttt{xbrlapi} sends to the BrlAPI server the ID of
the active window. The user can hence just switch between consoles and
windows as usual, and the braille device will always display the
appropriate application's output.

The notion of focus is even nested by having applications actually send
a list of integers: an application running on VT 2 will send a 1-element
list just holding $2$, while an application running in a window of an X
server that is running on VT 7 will actually send a 2-element list holding
$7$ and the window ID.  The BrlAPI server just has to first find out
which VT is active, and if this is VT 7, find out which window of the X
server is active.

Finally, as was mentioned in Section \ref{concurrency}, some specialized
applications like file transfer tools may have to take full control
of the braille device.
With our client-server approach, we
check that only one application at a time requests such
control, and disable all braille driver handling except the basic
low-level device protocol operations like splitting data coming from
the device into packets (if the device protocol is packet-oriented).
The application can then send raw data to the braille driver, which in
turn sends it to the device.  Symmetrically, data that the
braille driver receives from the device is passed back to the
application.  This way, applications can easily implement file
transfer protocols for instance.

Furthermore, applications can even switch to the ``suspended
mode'', in which the BrlAPI server keeps its braille driver shut down,
so that the application may open the device itself and thus be free to
totally control it.

The XVI \cite{xvi} client-server design is similar to ours, but though
it permits several clients to connect concurrently, it does not perform
an automatic switching between them.  It also does not allow applications
to take full control of the braille device.

\subsection{A Simple Interface for Programmers} \label{interface}

In order to achieve the various actions explained above, applications actually
switch between the following modes.

\begin{description}
\item[Tty mode:] The application may issue \verb+write+ requests for
displaying text on the braille display, and may receive \verb+key+ events.
\item[Raw mode:] The application directly communicates with the braille
device through the BrlAPI connection.
\item[Suspended mode:] The BrlAPI server keeps its device driver shut
down, letting the application directly open the device itself.
\end{description}

For instance, the following simplified code snippet connects to the
server, writes a prompt and waits for a key:

\begin{verbatim}
   brlapi_openConnection();
   brlapi_enterTtyMode();
   brlapi_writeText("Press any key");
   brlapi_readKey(&code);
   brlapi_leaveTtyMode();
   brlapi_closeConnection();
\end{verbatim}

\subsection{Protection} \label{protection_handle}

BrlAPI includes an authorisation mechanism used by the server to
decide whether a client is allowed to connect or not. The precise
authorisation procedures are left unspecified, so that various
authorisation procedures can be implemented according to what the
operating system supports (\verb+SO_PEERCRED+, \verb+getpeereid+, etc.).
When a client connects, the server advertises the list of supported
authorisation mechanisms. The client uses one of them, and if the
authorisation procedure succeeds, the connection is accepted. Otherwise, the
client may try a different mechanism.

\section{Implementation} \label{implementation}

BRLTTY \cite{brltty} is a portable screen reader: it works on
Linux, all BSD flavors (including MacOS X), Solaris, OSF, HP-UX, QNX,
Windows, Hurd, DOS. It also supports a wide range of braille
terminals (around 50 models).  The BrlAPI framework was hence
implemented within BRLTTY.  A BrlAPI server is implemented within the
screen reader in an independent thread,
and a library is provided so that applications can easily act as BrlAPI
clients.  Bindings for this library have been written for Python, Java,
TCL and Common lisp.  Bindings for some other languages (Objective Caml,
Perl and Haskell) are currently in development.

When both clients and server are running on the same machine, either a Unix
local socket or a Windows named pipe is used to establish the connection.
In some cases (just like with
X-window applications), clients may have to run on another machine. In
this case, a TCP/IP connection is used instead. A range of ports has
been reserved at IANA \cite{iana} for this purpose. In order to reduce the
risks inherent in text parsing and to keep the server simple, we use a
binary protocol over this connection. No encryption is performed, since just like X, it would be
easy to encrypt the connections using external tools like ssh or
IP-layer encryption.

Finally, the communication protocol has been designed to minimize the
number of packets to exchange between client and server in situations
where performance matters. For instance, write requests emitted by a client
require no acknowledgement from the server. If the write fails for some reason
(e.g. inconsistent data are sent), an error will be reported
{\em asynchronously}.

\section{BrlAPI Clients} \label{results}

The BrlAPI framework is already widely used on Linux systems as a
means for screen readers to drive braille devices. This
includes X-window readers like Gnopernicus from Baum \cite{gnopernicus},
Orca from Sun \cite{orca} and LSR from IBM \cite{lsr}, as well as
the Emacs reader speechd-el \cite{speechd-el}. This permits
not only to avoid having to re-implement device drivers, but also to
cooperate harmoniously with the Linux text console: BrlAPI allows
switching between the BRLTTY reading of Linux text consoles and the reading
of X or Emacs Windows.  The gnome-braille \cite{gnome-braille} braille
translation library also uses BrlAPI as one of its output devices.

Some file transfer applications have been developed and are now used
on a daily basis: \verb+vbtp+ for VisioBraille devices, \verb+trf+ for
EuroBraille devices, and soon \verb+htcom+ for HandyTech devices.

Also, the \verb!bless! client is a little braille-enabled replacement
for the well-known \verb!less! Unix command. Currently, it lets visually
impaired persons read documents without having to scroll the screen,
and allows them to save their current position in a document in order to
come back to it quickly later.

Finally, it is worth mentioning that the Accessibility Free Standards
Group \cite{a11y} is even considering adopting BrlAPI as its standard
braille I/O sharing component.

\section{Conclusion} \label{conclusion}

The large diversity of braille terminals makes it especially difficult to design
a generic communication interface which also allows the use of
advanced capabilities provided by each terminal.
Allowing several applications to access a braille display concurrently,
and in a way that protects the display from incorrect use are two other
important problems. BrlAPI solves these problems by relying on a
client-server approach: each application connects to a server, which is
responsible for abstracting away the details of each device and decides
which application should, at any moment, be allowed to communicate
with the device. That way, the end user doesn't have to care at all: she
can just switch between consoles and windows as usual, the braille
device will always display the appropriate information.
An implementation of the BrlAPI server has been
integrated to the BRLTTY screen reader, and a library has been provided
to help client applications communicate with braille devices through
BrlAPI. Currently, BrlAPI is used by several screen readers for graphical
environments.

It has been suggested that instead of using a client-server
approach, we could instead define a TTY terminfo entry, and applications
would not have to be modified: they would simply notice that they are running in a
very small terminal (compared to the usual 80x25 ones).  However, since
applications have only one standard output, that would also mean that sighted
users would be restricted to this output.

Although we are fully convinced that BrlAPI is a mature project, we also
think that a number of things remain to be done. For a start,
many users and developers would like to be able to spatially share a braille
window.
Instead of using the whole display as is the case for
now, applications would be given the control of only one region of the
display, leaving the other regions available for other applications. This
would not only be interesting for multi-line displays; it would also be
useful on long single-line displays, to reserve the rightmost part for displaying the current time on those braille terminals that do
not include status cells.

It would also be useful to write a detachable multiplexer. It would
for instance be run along with the \verb+screen+ program and allow the running
of applications in a ``detached'' session and let the user reattach the
session, including the BrlAPI applications.

Furthermore, we would like to let applications configure and request
the properties of the BrlAPI server.  These would of course include
the name of the currently loaded braille driver, the size of its braille
display and the current braille table (which defines the mapping
between characters and their braille representations), but also the
cursor blinking rate and shape, the braille contraction style, etc.

Yet another capability we may add to BrlAPI is support for
graphical braille displays. Indeed, in Section~\ref{variety}, we
asserted that braille displays are divided into eight-pin cells. However,
this is not always true as some terminals with a matrix of pins
have been designed, their goal being to display figures and graphical
information. Although this kind of terminal is rare, the situation
may change in the future, hence the need for BrlAPI to support them too.

Eventually, we would like to integrate the BrlAPI protocol directly into
X and TTY streams. That would solve a lot of focus and authorisation
issues.

\bibliographystyle{plain}
\bibliography{paper}

\begin{thebibliography}{10}

\bibitem{a11y}
{Free} {Standards} {Group} -- {Accessibility} {Workgroup}.
\newblock \url{http://www.a11y.org/}.

\bibitem{tim}
Dominique Archambault, Dominique Burger, and Sébastien Sablé.
\newblock The {TIM} project: Tactile interactive multimedia computer games for
  blind and visually impaired children.
\newblock In {\em AAATE 2001}, Ljubljana, September 2001.

\bibitem{gnopernicus}
Baum.
\newblock Gnopernicus.
\newblock \url{http://www.baum.ro/gnopernicus.html}.

\bibitem{xvi}
{BEAM}.
\newblock {XVI}.
\newblock \url{http://portal.beam.ltd.uk/xvil/sbserver_api.html}.

\bibitem{iana}
Internet~Corporation for Assigned~Names and Numbers ({ICANN}).
\newblock Internet assigned numbers authority ({IANA}).
\newblock \url{http://www.iana.org/assignments/port-numbers}.

\bibitem{speechd}
Free(b)soft.
\newblock {Speech} {Dispatcher}.
\newblock \url{http://www.freebsoft.org/speechd}.

\bibitem{speechd-el}
Free(b)soft.
\newblock Speechd-el.
\newblock \url{http://www.freebsoft.org/speechd-el}.

\bibitem{lsr}
IBM.
\newblock {Linux} {Screen} {Reader}.
\newblock \url{http://www.alphaworks.ibm.com/tech/lsr},
  \url{http://live.gnome.org/LSR}.

\bibitem{brltty}
Dave Mielke.
\newblock {BRLTTY}.
\newblock \url{http://mielke.cc/brltty/}.

\bibitem{gnome-braille}
The~{GNOME} Project.
\newblock {Gnome} {Braille}.
\newblock \url{http://cvs.gnome.org/viewcvs/gnome-braille/}.

\bibitem{orca}
The~{GNOME} Project.
\newblock Orca.
\newblock \url{http://live.gnome.org/Orca}.

\bibitem{libbraille}
Sébastien Sablé and Dominique Archambault.
\newblock Libbraille: A portable library to easily access braille displays.
\newblock In {\em International Conference on Computers Helping People with
  Special Needs (ICCHP)}, Linz, Austria, July 2002. Springer.

\end{thebibliography}
\end{document}